\documentstyle[multicol,aps,epsf,here]{revtex} 
\tighten

\newcommand{\PostScript}[7]{
\begin{figure}[H]
\begin{center} 
\leavevmode
\epsfysize=#1cm
\vspace{#2cm}
\epsfbox{#3}
\par
\parbox{#5cm}{
\vspace{#4cm}
\caption[figure]{\renewcommand{\baselinestretch}{1} \small \normalsize #6}
\label{#7}}
\end{center}
\end{figure}
}

\newcommand{\ie}{{\em i.e.} }

\newcommand{\noi}{\noindent}
\newcommand{\etal}{{\em et al.\ }}

\newcommand{\p}{\partial}
\newcommand{\pa}[1]{\partial_{#1}}
\newcommand{\pref}[1]{(\ref{#1})}
\newcommand{\zs}{(z_1,z_2,...,z_N)}

\newcommand{\zb}{\bar z}
\newcommand{\zt}{\tilde z}
\newcommand{\etat}{\tilde \eta}
\newcommand{\jas}{\prod_{i<j} (z_i - z_j)}

\newcommand {\be}[1]{
      \begin{eqnarray} \mbox{$\label{#1}$}}
\newcommand{\ee}{\end{eqnarray}}

\begin{document}

\title{Bose condensates at high angular momenta}

\author{S. Viefers$^{\$ , \dagger}$, T.H. Hansson$^*$ 
  and S.M. Reimann$^{\$}$}      

\address{$^\$$Department of Physics, University of Jyv\"askyl\"a,
Box 35, FIN-40351 Jyv\"askyl\"a, Finland}

\address{$^*$Department of Physics, University of Stockholm, Box 6730,
S-11385 Stockholm, Sweden }

\address{$^{\dagger}$NORDITA, Blegdamsvej 17, DK-2100 Copenhagen, Denmark}

\maketitle

\begin{abstract}
We exploit the analogy with the Quantum Hall (QH) 
system to study weakly interacting
bosons
in a harmonic trap.
For a $\delta$-function interaction potential the ``yrast'' states
with $L\ge N(N-1)$ are degenerate, and we show how this can
be understood in terms of Haldane exclusion statistics.
We present spectra for 4 and 8 particles obtained by numerical and
algebraic methods, and demonstrate how a more general hard-core potential lifts
the degeneracies on the yrast line. The exact wavefunctions for $N=4$ are
compared with trial states constructed from composite fermions (CF), and the
possibility of using CF-states to study the low $L$ region 
at high $N$ is discussed.
\\
%
\end{abstract}
\begin{multicols}{2}
\narrowtext

The close relation between high angular momentum states of a 
condensate of {\em weakly} interacting hard core bosons 
\cite{butts1,mottelson1,kavoulakis1}
and the Quantum Hall 
(QH) effect was recently pointed 
out~\cite{wilkin1,wilkin2,cooper1}. 
The essential observation is that the weak interaction limit 
allows for a two-dimensional
description of the boson system in terms of lowest Landau level (LLL)
wave functions~\cite{wilkin1} -- just as for a QH system, and they are  
both described by wave functions containing
powers of the Laughlin-Jastrow factor $\prod_{i<j} (z_i - z_j)$, where 
$z_i$ is the complex coordinate of the $i^{th}$ particle. As pointed out 
by Cooper and Wilkin~\cite{cooper1}, the two systems can in fact be 
mapped onto each other 
by a standard Leinaas-Myrheim transformation~\cite{leinaas1}, 
attaching an odd number of units of 
statistical flux to each particle. This enables us to use both intuition and 
techniques from the QH system to study rotating Bose condensates. 
In this paper we shall mainly discuss the angular momentum region 
$N(N-1)\le L\le 2N(N-1)$, where the ground state corresponding to a 
$\delta$-function two-body interaction is degenerate.
We  explicitly show
how these degeneracies can be understood via a mapping to a system of
free anyons in the LLL, and then show how the degeneracy is broken
by a more general short range potential containing derivatives of 
delta functions.
We also make a detailed comparison between 
algebraically calculated exact wave functions and trial wave functions 
formed from so called compact states of composite fermions, a 
construction orginally due to Jain and Kawamura~\cite{jain2}.
In particular, we will emphasize the importance of a certain class of 
wavefunctions where the polynomial part is translationally invariant. \\
Although for computational reasons we have results only for few 
particles, $N=4,6$ and 8, it is clear that some of our results, like 
the degeneracy structure of the yrast line, hold for any $N$. 
We also believe that many features of the results for the CF 
wave functions will generalize to higher $N$ . \\
Since the flux attachment changes the angular momentum by $mN(N-1)/2$, 
where $m$ 
is the number of fluxes attached, the boson - fermion mapping would apparently 
only be useful for studying  angular momenta that are out of reach of 
present experiments\cite{matthews1} 
(which are limited to the strong interaction
regime and $L\sim N$). However, there are some 
indications that fermionic techniqes could be useful for $L$ as low as $N$, 
\ie for the  so called single vortex state. Although we will 
 return to this point at the end of the
paper, we shall for now, without any further apologies, consider 
the theoretical 
problem of understanding
the region $N(N-1) \leq L \leq 2N(N-1)$ of the yrast line.

The simplest model for a hard-core interaction is a delta function potential.
We thus consider a model of $N$ interacting spinless bosons in a 
harmonic trap of strength $\omega$.
In the limit of weak interaction, this may
be rewritten \cite{wilkin1} as a two-dimensional lowest Landau level (LLL) 
problem in the effective
``magnetic'' field $B_{eff}=2m\omega$ with the Hamiltonian taking
the form
\be{Ham2}
H = \omega L + g \sum_{i<j}\delta^2({\bf r}_i-{\bf r}_j) \   
\ee
($\hbar=1$),
where $L\equiv\sum_i l_i = L_z$ is the total angular momentum. 
The single-particle states spanning our Hilbert space are
$\eta_{0,l}=(2^{l+1} \pi l!)^{-1/2} z^l \exp(-\zb z/4)$
with $z = \sqrt{2m\omega}(x+iy)$. 

In Figs.~\ref{f1} and~\ref{f2}  we show the interaction energy, in 
units of  $g/4\pi$ as a function of the total angular momentum $L$
for $N=4$, $L\le 20$  and $N=8$, $30 \le L\le 56$, respectively.
The many-body states are obtained from Lanczos diagonalization
suitable for large and relatively sparse matrices~\cite{arpack}.
The Fock space is spanned by single-particle states that are characterized 
only by the positive quantum numbers $l$, where $0\le l\le L$. 
(Similar exact diagonalization studies have recently been
reported in \cite{cooper1} and \cite{bertsch1}.)
We note the following properties in the many-body spectra:
Since increasing the angular momentum spreads out the 
particles in space,
the yrast energy, \ie the lowest possible interaction energy for
given $L$, decreases with increasing $L$. For each state in the spectrum,
there exists a set of ``daughter states'' with higher values of $L$,
having exactly the same (interaction) energy as the original state.
These daughters  are simply center-of-mass excitations of the
original states, thus having the same many-body correlations\cite{trugman1}.

\PostScript{9}{0}{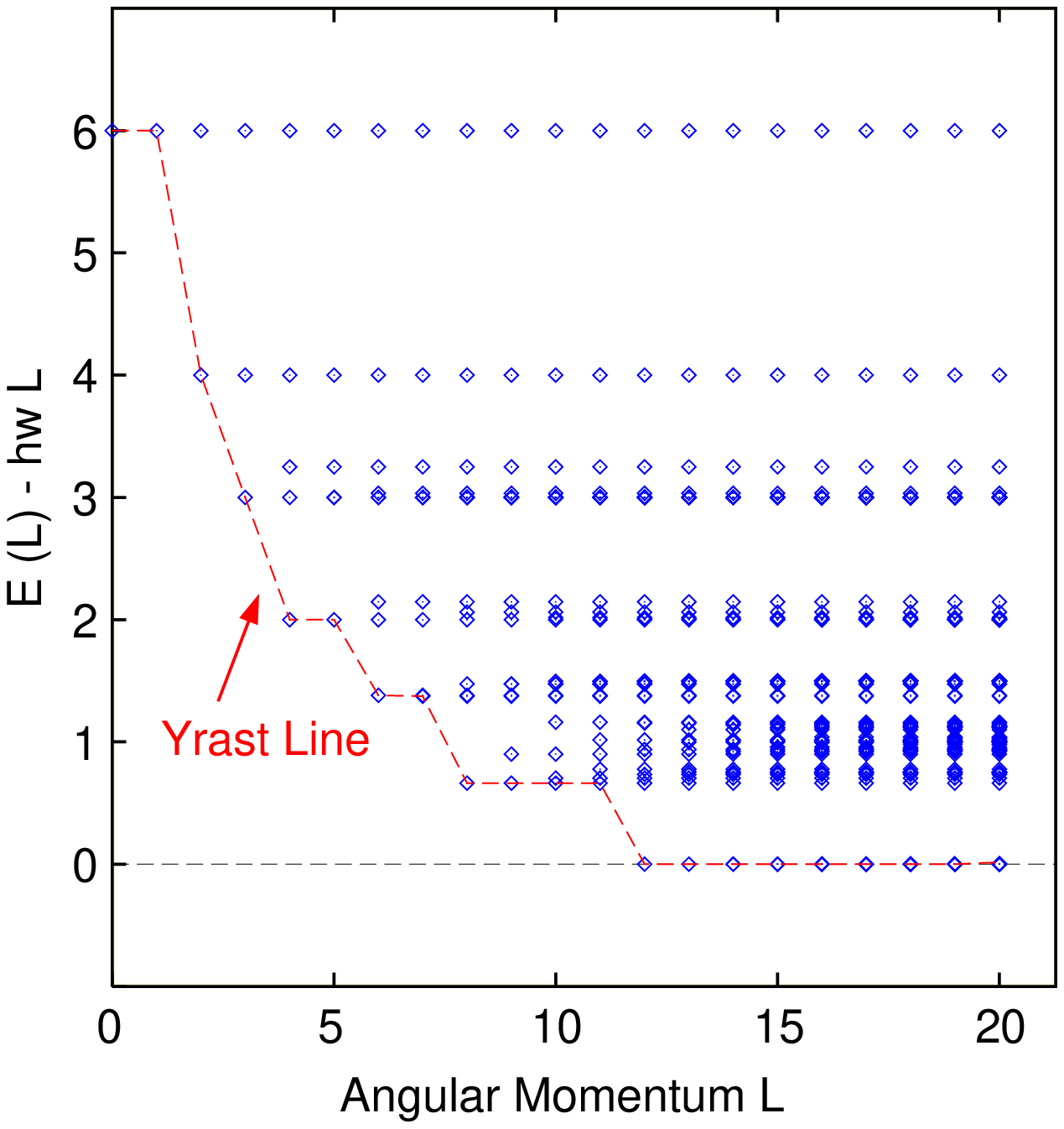}{0.5}{14}{\normalsize Many-body spectra of 
$N=4$ weakly interacting bosons in a harmonic trap for $L \leq 20$. The dashed
line connects the yrast states}
{f1}
\PostScript{6.4}{0}{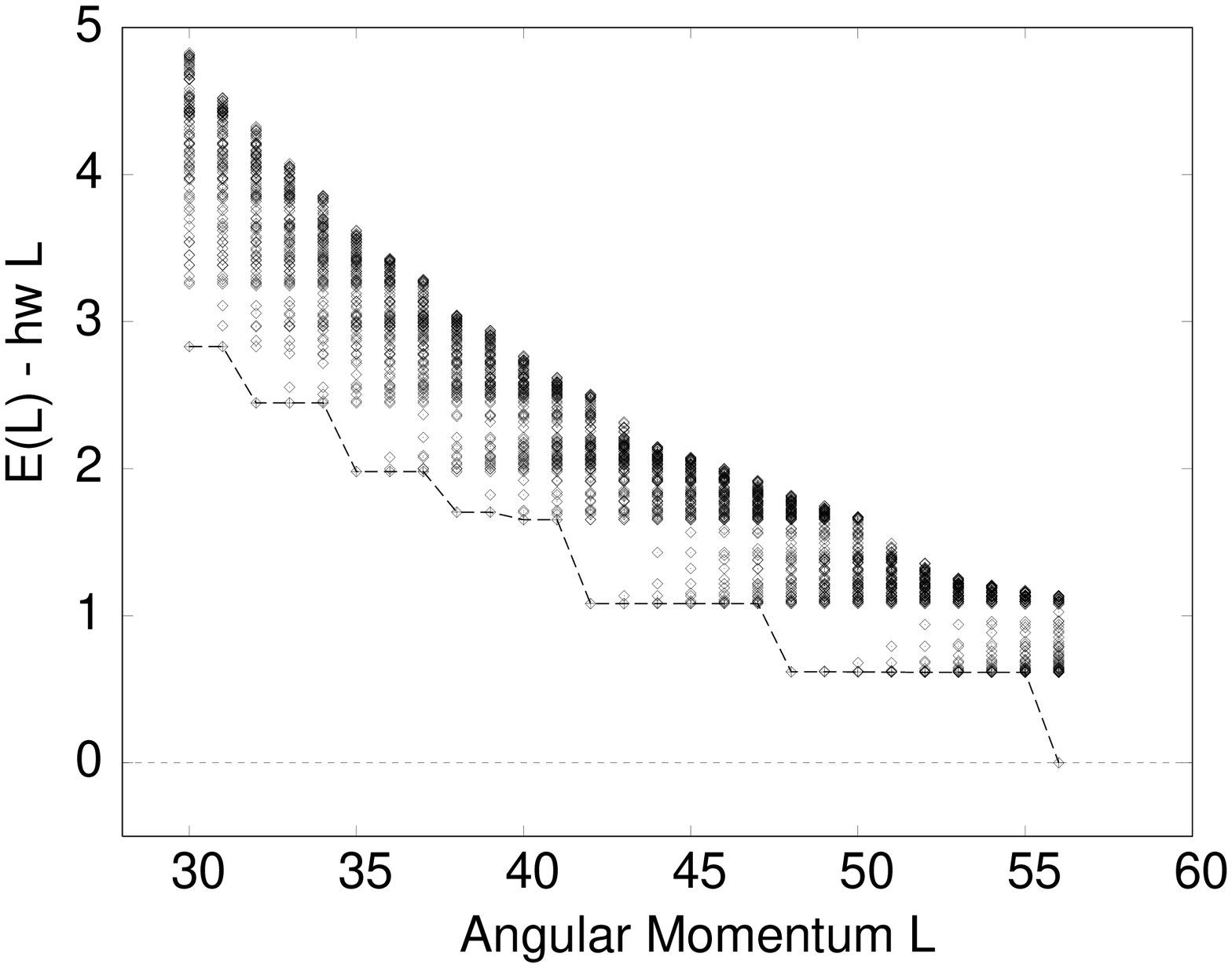}{0.5}{14}{\normalsize As Fig.~\ref{f1}, 
but for $N=8$ and $30 \le L \leq 56$, showing only the 100 lowest
eigenvalues}
{f2}
For $L\ge N(N-1)$ there are  zero energy states, which 
can be understood by noting that any wavefunction of the form 
\be{jasfor}
\psi\zs  = \jas^2 S\zs \ \  \ ,
\ee 
where $S\zs$ is a symmetric polynomial in the $z_i$:s,
has zero interaction energy. 
Since the factor $\jas^2$ contributes an angular momentum
$L_0=N(N-1)$, states of the type \pref{jasfor} exist for $L\geq L_0$.

At $L=N(N-1)$ there is a unique state with zero interaction energy
corresponding to $S\zs=1$, 
while the 
states at higher $L$ typically are degenerate. The systematics of these 
degeneracies can be understood by a mapping to anyons in the lowest
Landau level. 
The essential observation is that the wave functions \pref{jasfor} 
describe anyons in the LLL
with statistics parameter $\alpha=2$\cite{hansson1}
(in general, the statistics parameter is given by
the exponent of the Jastrow factor). 
It is 
known\cite{hansson1,isakov1,LLLFES} that anyons in the LLL obey Haldane's
fractional exclusion statistics (FES)\cite{haldane1}, and
following Ref.\cite{hansson1}, one can 
use this knowledge to construct the allowed many-body states for
given $N$ and $L$ as angular momentum excitations of the
Laughlin-like state at $L=N(N-1)$. According to the definition of
FES, each particle in the system blocks $\alpha=2$ single-particle
states, and many-particle states with total angular
momentum $L$ are constructed by occupying single-particle states, with a 
minimum distance of $\alpha=2$ between each pair of occupied levels
(for an example, see Fig.~\ref{f3}). The number of allowed configurations 
then gives the degeneracy of the state for a given $L$.

 \PostScript{4}{0}{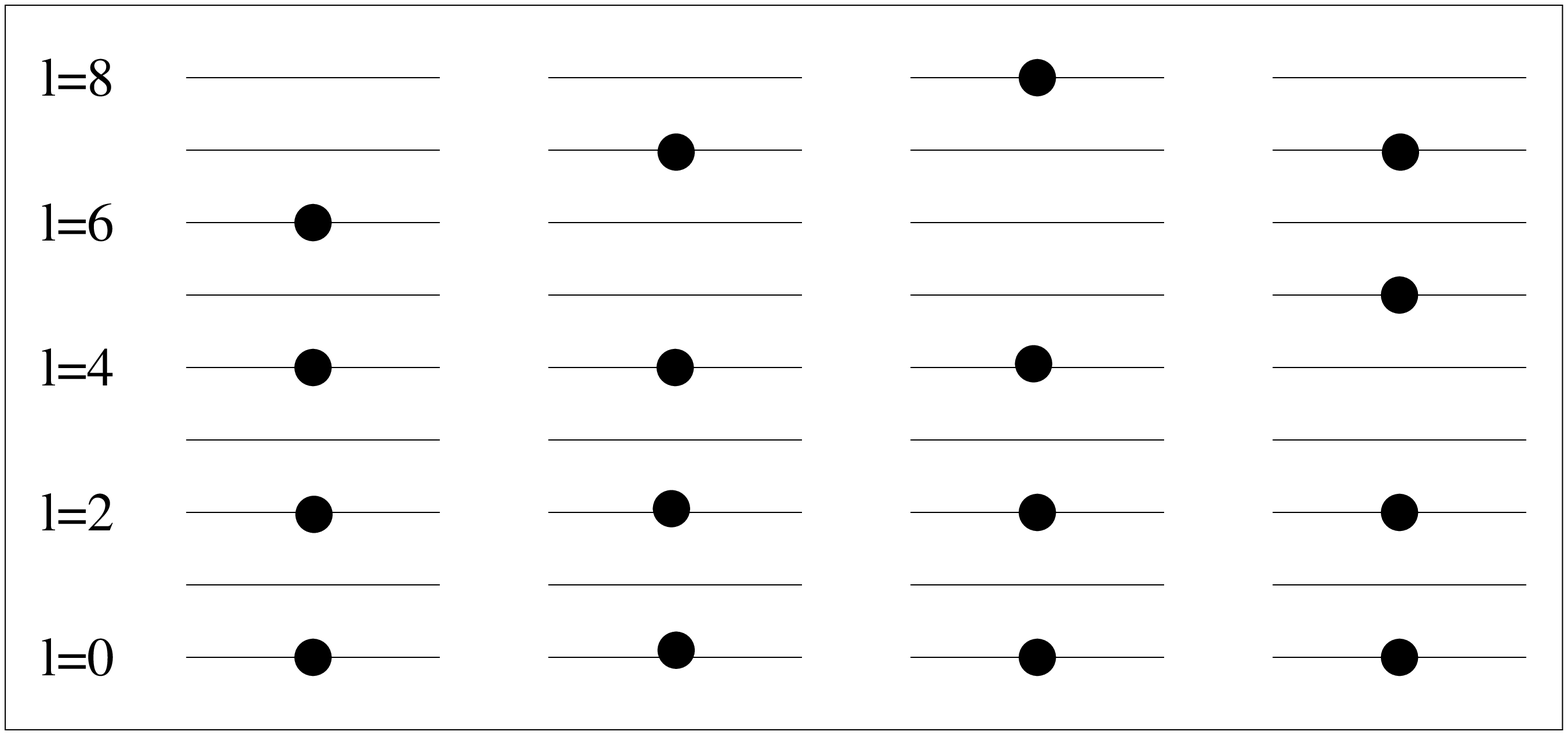}{0.5}{14}{\normalsize FES construction of
 yrast states for $N=4$, $L=12(=L_0)$, 13 and 14. The $L=12$ and
 $L=13$ states are non-degenerate, whereas $L=14$ has degeneracy 2.}
 {f3}

  Table 1 shows the degeneracies of some of the states on the $N=4$ yrast
  line, as obtained from the anyon mapping, and they  are in exact 
  agreement with our  numerical results.

\bigskip

  \begin{table}[htb]
  \begin{center}	  
    \begin{tabular}{rrrrrrrrrr}                  
    $L$ & 12 & 13 & 14 & 15 & 16 & 17 & 18 & 19 & 20      \\  \hline 
    $d$ & 1  &  1 &  2 &  3 &  5 &  6 &  9 & 11 & 15        
     \end{tabular}

\bigskip

  \caption{Degeneracy $d$ of the lowest $L$-excitations above
  the Laughlin state for $N=4$.}
  \label{tab1}
  \end{center}
  \end{table}
This construction implicitly uses that 
all the eigenstates on
the yrast line for $L > N(N-1)$ contain the Jastrow factor $\jas^2$, 
so 
the degeneracies can also  be found  as the number
of ways one can distribute $M=L-L_0$ units of angular momentum
among $N$ particles.

For a more general hard core potential, the $E=0$ states on the yrast line 
above $L=N(N-1)$ are no longer degenerate. To demonstrate this point and
study how the degeneracy is broken, we add a potential of the form
\be{ddpot}
V=c_1 \nabla^2\delta^2(z-z') + c_2 \nabla^4 \delta^2(z-z') ,
\ee
that was originally used by 
Trugman and Kivelson\cite{trugman1} 
in the context of the fractional QH effect. 
The term $\sim\nabla^2\delta^2(z-z')$ does not 
contribute to the interaction energy for fully symmetric states,
whereas the $\nabla^4\delta^2(z-z')$ term gives small corrections to the
spectra in Figs.~\ref{f1} and~\ref{f2} (at the percent level for
the parameters used in the inset of Fig.~\ref{f5}, 
where we show regularized forms
of the potentials~\pref{Ham2} and~\pref{ddpot}).


We have examined how the potential \pref{ddpot}
splits up the degeneracies of the zero interaction energy 
yrast states, by exact 
algebraic diagonalization, using computer algebra. 
Here we have directly used the form 
\pref{jasfor}, and systematically exploited that for a
given $L$, all states corresponding to center-of-mass excitations
of lower $L$-eigenstates are orthogonal to the subspace of ``new''
states. The latter subspace consists of translation invariant (TI)
polynomials, \ie functions invariant under a simultaneous, constant
shift $z_i \rightarrow z_i+a$ of all the coordinates\cite{trugman1}.
Following Trugman and Kivelson\cite{trugman1}, we have used a basis
constructed from elementary symmetric functions $s_n$. 
For given $N$ and
$L$, the basis consists of all possible combinations
\be{basis}
|k_1 k_2 ... k_n \rangle \equiv s_1^{k_1}(z_i) s_2^{k_2}(\zt_i) ...
                                s_N^{k_N}(\zt_N) \, \jas^2,
\ee
such that $\sum_{n=1}^N n k_n = L-L_0$. Note that,
for $n\geq 2$, we have introduced the
new variables $\zt_i \equiv z_i - z_c$, with $z_c$ the center-of-mass 
coordinate $z_c = (\sum z_i)/N$.
The basis states spanning the TI subspace
are identified as those with $k_1=0$.
The diagonalization is thus performed within this 
subspace only, which reduces the matrix dimension substantially.
The resulting spectrum for $N=4$, with the coefficient $c_2$ in
\pref{ddpot} set equal to 1, 
is shown in Fig.~\ref{f5}.
We notice the close similarity between Fig.~\ref{f1} and \ref{f5}. 
In both cases, the yrast line passes through the same number of
steps, with the same step lengths, as $L$ is increased by $N(N-1)$,
to the point where the yrast energy becomes zero.
At the point $L=2N(N-1)$, the zero-energy yrast state is again
non-degenerate and of the Laughlin type, \ie $\jas^4$, whereas
the subsequent yrast states have degeneracies corresponding to
FES with statistics parameter $\alpha = 4$. Including even higher derivative
terms in the repulsive potential would subsequently split up the
higher regions of the yrast line. 
Finally, note that the yrast states corresponding to cusps, \ie states 
that are followed by a ``plateau'' in the yrast line,
are always in the TI subspace. 
\PostScript{7}{0}{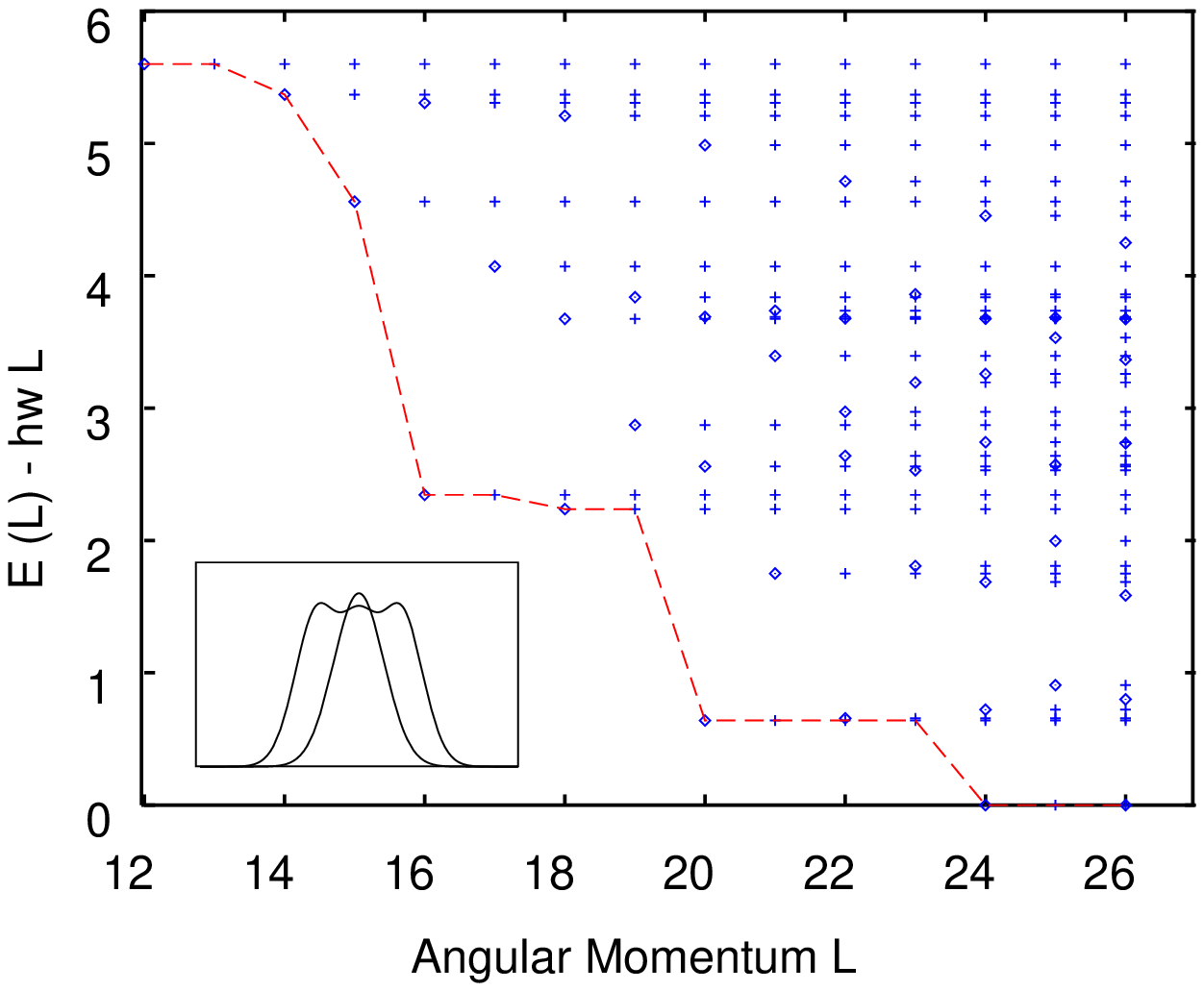}{0.5}{14}{\normalsize Yrast spectrum for $N=4$ 
                 and $12 \leq L \leq 26$, for a repulsive potential
                 $\nabla^4\delta^2(z-z')$. Translation
                 invariant eigenstates are denoted by diamonds, 
                 whereas the crosses denote center-of-mass excitations.
{\it Inset:} Delta function
potential (regularized as a gaussian) and a potential including
the terms \pref{ddpot}, with $c_{1}=0.023$ and $c_{2}=0.00008$.
We see that the latter more closely resembles a hard sphere 
potential.}
{f5}

The algebraic diagonalization used here is limited in
practice to smaller particle numbers and angular momenta than
the numerical scheme used in Figs.~\ref{f1} and \ref{f2}. 
However, the present
approach has the advantage that it provides explicit, analytic
expressions for the eigenfunctions, in terms of symmetric
polynomials. This gives some additional insight into the structure
of the yrast states, and in particular the region below the
single vortex in the case of a pure delta function interaction. Bertsch 
and Papenbrock \cite{bertsch1} recently proposed and numerically
tested the following form for the yrast states at $2 \leq L \leq N$,
\be{bwf}
\psi_L(z_i) &=& \sum_{p_1 < p_2 < ... < p_L}
            (z_{p_1} - z_c)(z_{p_2}-z_c)...(z_{p_L}-z_c) \ .
\ee 
This is just the symmetric polynomial $s_L(\zt_i)$, 
\ie the state
$|0...1..0\rangle$ (with the 1 in the $L$th place), in the notation of
\pref{basis} (without the Jastrow factor in the present case
of a pure delta function interaction). This state is a basis state
in the TI subspace for all $2 \leq L \leq N$.
In the cases where this is the {\em only} basis state
($L=2,3$ for $N\ge 3$), it is thus 
obvious that \pref{bwf} is exact.
Furthermore, performing the algebraic diagonalization up to $L=N$
for $N=4$ and $N=6$, we have confirmed that even when the
translation invariant subspace is spanned by more than one basis
vector, the basis state \pref{bwf} is always an exact eigenstate.
 
\bigskip
We now turn to a study of a class of wave functions that can be constructed
in analogy with the so-called Jain states for the fractional QHE\cite{jain1}. 
The main 
idea is to map the strongly interacting LLL bosons 
to weakly interacting composite fermions by attaching
an odd number $m$ of flux quanta to each particle. 
Trial wavefunctions with angular momentum $L$ are thus 
constructed as non-interacting
fermionic wavefunctions with angular momentum
$L- m/2 \, N(N-1)$, multiplied by $m$ Jastrow factors and projected
onto the LLL, 
\be{jainwf}
\psi_L = {\cal P} \left( f_F(z_i, \zb_i) \, \jas^m  \right) \ \ \ \ .
\ee
Here, $f_F$ is a Slater determinant consisting of single-particle
wave functions $\eta_{n,l}(z,\zb) \propto z^l L_n^l(\zb z/2)$ where $n$
is the Landau level index ( $l\geq -n$), and $ L_n^l$ a generalized 
Laguerre polynomial. 

Originally used for the homogeneous states relevant for the fractional QHE, 
wavefunctions of the type \pref{jainwf} were later employed to describe 
inhomogeneous systems such as quantum dots\cite{jain2,rejaei1,rejaei2} and  
recently  by Cooper and Wilkin \cite{cooper1} to study the bosonic
yrast lines for up to 10 particles in the case of a pure delta 
function interaction.

In short, the LLL projection in \pref{jainwf} amounts to the replacement
$\zb_i \rightarrow 2\pa{i}$ in the polynomial part of the wave function.
However, there are several ways  of doing this in practice, and we shall 
compare the different projection methods when constructing trial
wave functions for the yrast states in Figs.~\ref{f1} and \ref{f5}.
The most straightforward way is to replace the $\zb$:s with derivatives 
in the final polynomial, obtained after multiplying out the 
Slater determinant and the Jastrow factors and moving all $\zb$:s
to the left. In practice, this method is applicable only for
small particle numbers, when the number of derivatives involved is
not too large. We shall refer to this method as ``Method I'' and apply
it in a few special cases for comparison.

Alternatively, noting that\cite{jain1}
 \be{minijastrow}
 \left|
 \begin{array}{ccc}
 \eta_{11} & \eta_{12}  & ..     \\
 \eta_{21} & \eta_{22}  & ..     \\
 ...               &     ...            & ..     \\
 \eta_{N1} & \eta_{N2}  & .. 
 \end{array}
 \right|
 \jas^{2p}   
 = 
 \left|
 \begin{array}{ccc}
 \etat_{11} & \etat_{12}  & ..     \\
 \etat_{21} & \etat_{22}  & ..     \\
 ...               &     ...            & ..     \\
 \etat_{N1} & \etat_{N2}  & .. 
 \end{array}
 \right| ,
 \ee
where 
$\eta_{ij}\equiv \eta_i(z_j,\zb_j)$
and
$\etat_{ij}\equiv \eta_i(z_j,\zb_j) \prod_{k\neq j}(z_j-z_k)^p$,
one can first absorb $2p$ Jastrow factors in the Slater
determinant, and then project entry by entry.
Since the wave function \pref{jainwf} contains an odd number $m$
of Jastrow factors, one finally has to
compensate by multiplying the resulting
polynomial by $\jas^{m-2p}$. We shall refer to this as Method II and
use it as follows; for a delta function interaction, 
where the wave function contains $m=1$ Jastrow factor, it is
appropriate to use $p=1$ (Method IIa).
In the case of the $\nabla^4\delta$-potential, the Jain construction
will involve absorbing $m=3$ flux quanta in the wave function,
and we shall compare projection with $p=1$ (Method IIa) and $p=2$
(Method IIb). (Note that Method IIa is in a sense trivial, since 
it relates   the wavefunctions  at $L$ and $L+N(N-1)$ by 
multiplication with two Jastrow factors.)

We have already stressed the significance of the  TI 
subspace - these are the states that determine the shape of
the yrast line. It is very appealing that there is a special set 
of the states \pref{jainwf} that are in this subspace, namely
the so called {\em compact states} \cite{jain2} 
which are characterized by having
the $n$:th Landau level occupied from $l_{n}= -n$ to $l_{n} =l_{n}^{max}$ 
without any ``holes''. In the context of the QH effect, the 
important property of the compact states is that they are homogeneous.
When describing  quantum dots using the non-interacting composite 
fermion model (NICFM), one can show that CF candidates for the 
cusp states must be compact\cite{jain2}, and the same line of
arguments was later used for bosons~\cite{rejaei1,cooper1}.
From the point of view  
of the wavefunctions, the importance of the compact states is that 
they are in the TI subspace, and it is rather remarkable that 
{\em for all L where a compact state can be constructed, the one 
with the lowest CF energy has a large overlap with the lowest
exact state in the TI subspace.} This is true, independent of whether
or not the state is a ground state, \ie corresponds to a cusp in 
the yrast line. In the following discussion, we shall only be
concerned with the wavefunctions, and will not discuss whether
or not the NICFM can give a good description of the energy spectrum,
a question that was already discussed in some 
detail~\cite{jain2,rejaei1,rejaei2}.

\noi
{\bf $\delta$-function potential:} 
In this case we have calculated the Jain wavefunctions \pref{jainwf} 
corresponding to all compact  states for $N=4$, $0\le L \le 12$, 
taking $m=1$ and using   projection method IIa ($p=1$). 
If there are two compact states with the same $L$, we use the one 
with the lowest CF effective energy. 
In table \ref{tab2} we show the overlap with 
the exact algebraic wavefunctions. A ``1'' indicates that the 
wavefunctions are identical. For comparison we have also included the 
overlaps with the wavefunctions corresponding to projection method I, 
as given by Cooper and Wilkin\cite{cooper1}. 
We  note that {\em all} the compact states have very large 
overlap with the exact eigenstates.
That the CF wavefunction is 
exact for $L=2,3$ is a simple consequence of the TI property and that 
there is only one state in the TI subspace at these $L$ values. That 
the $L=7$ state comes out exact is more surprising, and we have no good 
explanation for this. It would be interesting to pursue the exact 
diagonalization to higher $N$  in order to see if there are other 
non-trivial CF states that are exact. We also note that the 
direct projection (method I) does slightly better than method II.

\bigskip

\begin{table}
\begin{center}
\begin{tabular}{rrrrrrrrr} 
$L$          & 0 & 2 & 3 & 4     & 6     & 7 & 8     & 12      \\  \hline 
${\rm IIa} $ & 1 & 1 & 1 & 0.944 & 0.962 & 1 & 0.997 & 1     \\ 
${\rm I}  $  & 1 &   &   & 0.980 & 0.980 &   & 0.997 & 1     
\end{tabular}

\bigskip

\caption{Overlaps between trial and exact wave yrast functions
for $N=4$, $0 \leq L \leq 12$ and a pure delta function interaction, using
projection methods IIa and I (see text).}
\label{tab2}
\end{center}
\end{table}

\noi
{\bf $\nabla^4\delta$-function potential:} 
Here we  calculated the Jain wavefunctions \pref{jainwf} 
corresponding to all, lowest CF-energy, compact  states for 
$N=4$, $12\le L \le 24$ 
taking $m=3$ and using 
projection methods IIa ($p=1$) and IIb $(p=2$). For comparison, we also
used method I for three $L$ values. 
In table \ref{tab3} 
we show the overlap with the exact algebraic wavefunctions. 
We again note that {\em all} the compact states (except $L=17$
with projection method I) have very large 
overlap with the exact eigenstates, and that for $L=17, 19$, the CF 
state is  not the ground state. 
Comparing the different projection methods, we see that in many, but 
not all cases they give identical results. 
In particular, one should 
note that method IIb does not reproduce the exact wavefunction for 
$L=14,15$ where the TI subspace is non-degenerate. This means that 
that projection gives a wave function that is not in the space of 
states given by \pref{jasfor}. The same effect is even more striking
for $L=17$, where method IIb gives the exact
wave function, whereas direct projection (method I) gives
a rather poor overlap, indicating that a large component is not
in the subspace \pref{jasfor}.

\bigskip

 \begin{table}[H]
 \begin{center}
 \begin{tabular}{rrrrrrrrrr} 
       $L$   & 12    & 14    & 15    & 16    &$17^*$ & 18    &$19^*$ & 20    & 24     
 \\  \hline 
 ${\rm IIa}$ & 1     & 1     & 1     & 0.988 &       & 0.910 & 0.993 & 0.986 &  1  
 \\ 
 ${\rm IIb}$ & 0.938 & 0.910 & 0.988 & 0.990 & 1     & 0.910 & 0.993 & 0.986 &  1  
 \\ 
 ${\rm I}$   &       &       &       &       & 0.591 &       & 0.993 & 0.986 &  1  
 \end{tabular}

\bigskip

 \caption{Overlaps between trial and exact yrast wave functions
 for $N=4$, $12\leq L \leq 24$, using two different projection methods
 (see text). The asterisk indicates that the $L=17$ and $L=19$ 
 Jain states (which
 correspond to the lowest TI states) are
 not yrast states, but lie higher in energy than the CM excitation
 of the previous $L$ state. Note that projection method IIa gives zero
 for $L=17$.} 
 \label{tab3}
 \end{center}
 \end{table}
\bigskip
Finally, we comment on the possibility to use CF wavefunctions
in the experimentally more relevant case of low $L$, and in particular
for the single vortex state at $L=N$.
Rather surprisingly, the overlaps between the CF  and the exact
wavefunction
\pref{bwf} tend to get {\em larger} with increasing particle number, at
least
up to $N=10$.\cite{cooper1}
It thus seems worthwhile to use the CF approach to construct
trial wavefunctions for the single vortex for general $N$.
The single vortex CF state is in fact unique, if one demands that in
addition to being compact, it should also have minimal CF cyclotron
energy.
The relevant Slater determinant is formed from the single particle
states $\eta_{n,-n}$ for $n= N-2, N-1,...0$ and  $\eta_{01}$, and,
using projection method I, the resulting wavefunction takes  the
following rather compact form,
\be{sv}
\psi_{L=N}= \sum_{n=1}^N (-1)^n z_n
            \prod_{k<l; k,l\neq n}\left( \p_k - \p_l \right) \jas .
\ee
Although the $\sim N^{2}$ derivatives make it difficult to evaluate
this function for large $N$, it can easily be handled in integrals
of the form 
$\int\prod_{i}d^2 z_{i}\,\exp(-1/2\sum_{i}\zb_{i}z_{i})
f(\zb_{i}) \psi_{L=N} $ by partial integrations. Alternatively
one can use  projection Method II, where the wavefunction becomes a
determinant of linear combinations of elementary symmetric polynomials.
In both cases, it should be possible to compare with the
exact wavefunction \pref{bwf} using Monte Carlo methods.

\vskip 3mm
\noi
{\bf ACKNOWLEDGEMENT}:  
We are very grateful to B. Mottelson for introducing
us to the physics of rotating Bose condensates,
and for numerous fruitful discussions. We also wish
to thank M. Manninen, G. Kavoulakis and R.K. Bhaduri 
for discussions and M. Koskinen
for giving advice on the numerical work. 
This work was financially supported by the Academy of Finland,
the Swedish Natural Science Research Council,
the TMR programme of the European Community under contract 
ERBFMBICT972405, the ``Bayerische Staatsministerium f\"ur 
Wissenschaft, Forschung und Kunst'', and the NORDITA Nordic
project ``Confined electronic systems''.

\vskip 70mm
\eject

\end{multicols}
\end{document}